\documentclass[twocolumn,aps,prl,showpacs,raggedbottom,nobalancelastpage,amssymb,superscriptaddress]{revtex4-1}

\usepackage{graphicx}
\usepackage{amsmath}
\usepackage{amssymb}
\usepackage{bm}
\usepackage[usenames]{color}
\usepackage{amsfonts}
\usepackage{appendix}
\usepackage{dsfont}

\def \H{\mathcal{H}}
\def \Fst{1^\textrm{st}}
\def \Snd{2^\textrm{nd}}

\def \Tr{\textrm{Tr}}
\def \l{\lambda}
\def \xvec{\mathbf{x}}

\def \P{\bar{P}}

\begin{document}
\title{Four-Dimensional Quantum Hall Effect in a Two-Dimensional Quasicrystal}

\author{Yaacov E.~Kraus}
\affiliation{Department of Condensed Matter Physics, Weizmann Institute of Science, Rehovot 76100, Israel}
\author{Zohar Ringel}
\affiliation{Department of Condensed Matter Physics, Weizmann Institute of Science, Rehovot 76100, Israel}
\affiliation{Theoretical Physics, Oxford University, 1, Keble Road, Oxford OX1 3NP, United Kingdom}
\author{Oded Zilberberg}
\affiliation{Department of Condensed Matter Physics, Weizmann Institute of Science, Rehovot 76100, Israel}
\affiliation{Institute for Theoretical Physics, ETH Zurich, 8093 Z{\"u}rich, Switzerland}


\begin{abstract}
One-dimensional (1D) quasicrystals exhibit physical phenomena associated with the 2D integer
quantum Hall effect. Here, we transcend dimensions and show that a previously inaccessible phase of
matter --- the 4D integer quantum Hall effect --- can be incorporated in a 2D quasicrystal. Correspondingly,
our 2D model has a quantized charge-pump accommodated by an elaborate edge phenomena with
protected level crossings. We propose experiments to observe these 4D phenomena, and generalize our
results to a plethora of topologically equivalent quasicrystals. Thus, 2D quasicrystals may pave the way to
the experimental study of 4D physics.
\end{abstract}


\maketitle

The uprising field of topological phases of matter deals with systems of arbitrary dimension~\cite{RMP_TI,RMP_TI2}. In this paradigm,
each energy gap of a system is attributed an index, which is robust to continuous deformations. A nontrivial index is usually associated with interesting boundary phenomena, quantized response, and exotic quasiparticles. While nontrivial topological phases appear in any dimension \cite{LudwigDim,KitaevDim}, the physical manifestations are limited to 1D \cite{Kouwenhoven2012,Das:2012}, 2D \cite{Kiltzing1980,Konig2007} and 3D \cite{BiSe_Hasan}.

An example of an intriguing topological phase, which is seemingly out of reach, is the 4D generalization of the 2D integer quantum Hall effect (IQHE). In 2D, a uniform magnetic field creates Landau levels that are characterized by the $\Fst$ Chern number—the topological index that corresponds to the quantized Hall conductance~\cite{TKNN,Avron1}. In 4D, a uniform SU(2) Yang-Mills field results in generalized Landau levels~\cite{Yang78,IQHE4D}. These levels are characterized by the $\Snd$ Chern number -- a topological index which corresponds to a quantized non-linear response~\cite{IQHE4D,QiZhang}. Both the 2D and 4D IQHEs exhibit a variety of exotic strongly correlated phases when interactions are included \cite{IQHE4D,FQHE_review}. Hence, the 4D IQHE, as well as lattice models with non-vanishing $\Snd$ Chern numbers, constantly attract theoretical attention~\cite{QiZhang,LiZhang2012}, such as the recently found rich metal-insulator phase diagram~\cite{Edge2012}.

The physical properties of quasicrystals (QCs) -- non-periodic structures with long-range order -- can oftentimes be derived from
periodic models of higher dimensions~\cite{QC_Senechal,Rodriguez2008}. Using a novel dimensional extension, it was recently shown
that 1D QCs exhibit topological properties of the 2D IQHE~\cite{us}. The bulk energy spectrum of these 1D QCs is gapped, and each
gap is associated with a nontrivial $\Fst$ Chern number. This association relies on the fact that the long-range order harbors an additional
degree-of-freedom in the form of a shift of the quasiperiodic order. Accordingly, boundary states traverse the gaps as a function
of this shift. This property was observed in photonic QCs, and was utilized for an adiabatic pump of light~\cite{us}. Moreover,
upon a deformation between QCs with different $\Fst$ Chern numbers, the expected phase transition was experimentally observed~\cite{us4}.
Generalizations to other 1D symmetry classes, physical implementations, and QCs were discussed~\cite{citing_us2,us2,Xu:2012,Ganeshan:2013}.

In this Letter, we take a major step further, and present a 2D quasiperiodic model that exhibits topological properties of the 4D
IQHE. Each gap in its energy spectrum is characterized by a nontrivial $\Snd$ Chern number, which implies quantum phase transitions between
topologically distinct models. Furthermore, scanning of the shift parameters is accompanied by (i) quantized charge pumping with an
underlying 4D symmetry, and (ii) gap-traversing edge states with protected level crossings. Generalizations to other models and 2D QCs are discussed. We propose two experiments to measure the $\Snd$ Chern number via charge pumping, and, thus, make 4D physics experimentally accessible.

\begin{figure}
\centering
\includegraphics[width=\columnwidth]{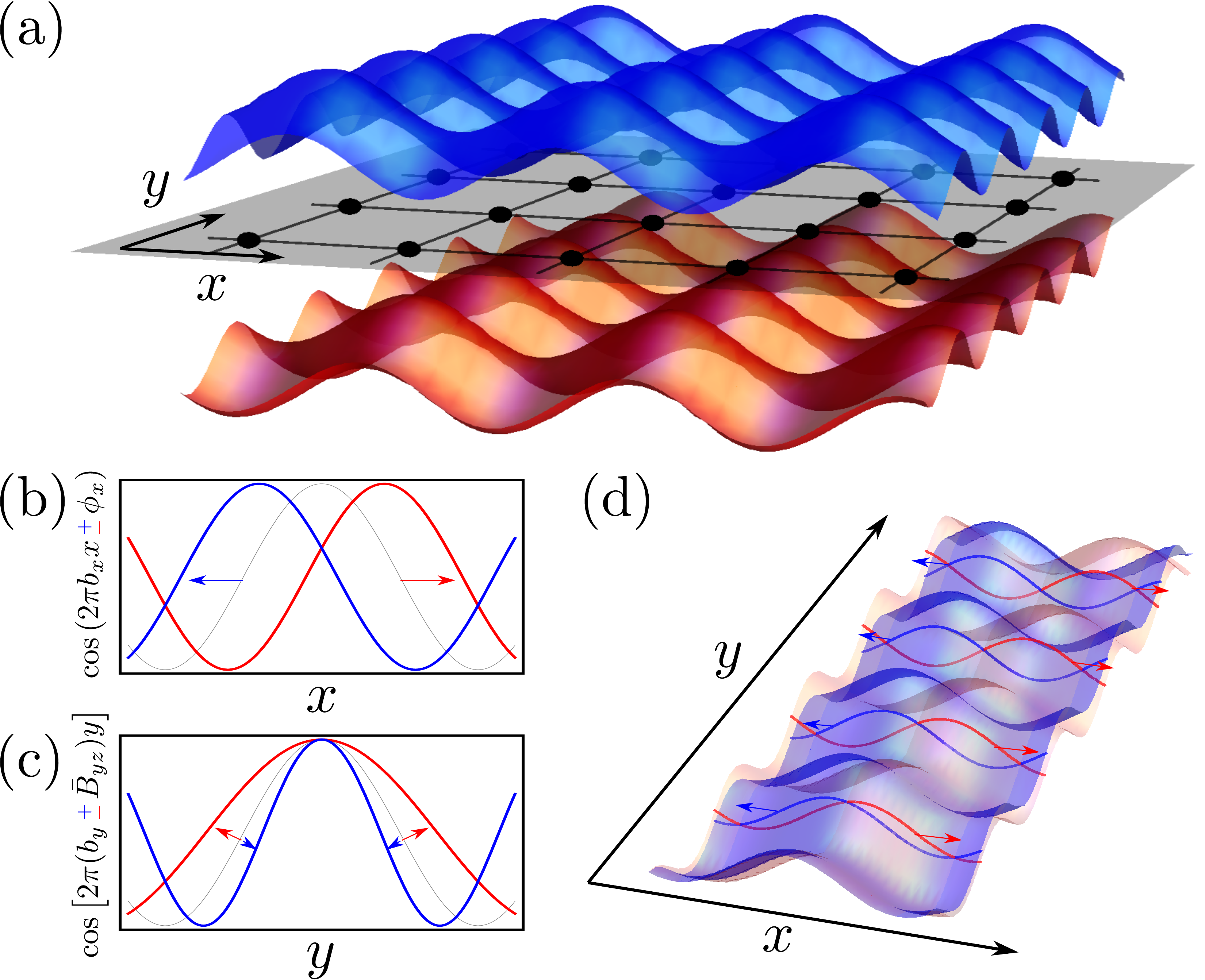}
\caption{\label{system_experiment} %
(a) An illustration of $H$ [cf.~Eq.~\eqref{Eq:H2D}]: a 2D square lattice with a cosine-modulated on-site potential.  The
potential is incommensurate with the lattice, and is $\sigma$-dependent. The state $\sigma=+$ experiences the upper (blue)
potential, and $\sigma=-$ experiences the lower (red). For clarity, the potentials are vertically-displaced. The dots mark the
underlying lattice sites. (b)-(d) A quantized charge pumping along the $x$ direction is achieved by scanning the shift parameters
$\phi_x$ and $\phi_y$ in the presence of the modulation modifications $\bar{B}_{yz}$ and $\bar{B}_{wy}$, respectively [cf.
Eqs.~\eqref{Eq:modification} and (6)]. The effects on the potentials of $\sigma=+$ (blue) and $\sigma=-$ (red) is illustrated
(the thin (gray) line denotes the unmodified reference potential): (b) $\phi_x$ (or equivalently $\phi_y$) shifts the
cosine-potentials in opposite directions for opposite $\sigma$. (c) $\bar{B}_{yz}$ makes the modulation frequency become
$\sigma$-dependent $b_y +\sigma \bar{B}_{yz}$. (d) $\bar{B}_{wy}$ shifts the cosine-potentials in opposite directions, but with
an increasing shift along the $y$ direction.}
\end{figure}

We study a 2D tight-binding model of particles that hop on a square lattice in the presence of a modulated on-site potential
\begin{align} \label{Eq:H2D}
    &H(\phi_x,\phi_y) = \sum_{x,y}\sum_{\sigma=\pm} c_{x,y,\sigma}^\dag \Big[ t_x c_{x+1,y,\sigma} + t_y c_{x,y+1,\sigma} + \text{H.c.} \nonumber\\
      & \resizebox{.9\hsize}{!}{$\displaystyle + \Big(\l_x \cos(\sigma 2\pi b_x x + \phi_x) + \l_y \cos(\sigma 2\pi b_y y + \phi_y)\Big) c_{x,y,\sigma} \Big].$}
\end{align}
Here $\sigma=\pm$ is an internal degree of freedom such as spin-$\frac{1}{2}$, photonic polarization, or atomic orbital,
$c_{x,y,\sigma}$ is the single-particle annihilation operator of a particle at site $(x,y)$ in state $\sigma$; $t_x,t_y$ are the
hopping amplitudes in the $x$ and $y$ directions; and $\l_x,\l_y$ are the amplitudes of the on-site potentials, which are
modulated along $x$ and $y$ with modulation frequencies $b_x,b_y$, respectively [cf.~Fig.~\ref{system_experiment}(a)]. Last,
the Hamiltonian depends on two shift parameters $\phi_x$ and $\phi_y$. We assume the modulation frequencies $b_x$ and $b_y$ to be
irrational, which makes the on-site modulations incommensurate with the lattice, and the model becomes quasiperiodic.

The spectrum of $H$ is gapped, and therefore, may exhibit nontrivial topological indices. Conventionally, the only apparent
topological index that can be associated with $H$ is the $\Fst$ Chern number~\cite{LudwigDim,KitaevDim}. However, since $H(\phi_x=\phi_y=0)$ is
time-reversal symmetric, its $\Fst$ Chern number vanishes, and $H$ is seemingly trivial.

Strikingly, $H$ is attributed a nontrivial $\Snd$ Chern number, which, by definition, characterizes 4D systems. In order to obtain this result,
let us consider $H$ on a toroidal geometry~\cite{torus}. We introduce twisted boundary conditions along the $x$ and $y$
directions, parameterized by $\theta_x$ and $\theta_y$, respectively. For a given gap and given $\phi_\mu \equiv
(\phi_x,\phi_y,\theta_x,\theta_y)$, we denote by $P(\phi_\mu)$ the projection matrix on all the eigenstates of $H(\phi_\mu)$ with
energies below this gap. We can now define,
\begin{align} \label{Berry_P}
\mathcal{C}(\phi_\mu) = \sum_{\alpha\beta\gamma\delta}\frac{\epsilon_{\alpha\beta\gamma\delta}}{-8\pi^2}
            \Tr \Bigg( P \frac{\partial P}{\partial \phi_\alpha} \frac{\partial}{\partial \phi_\beta}
                       P \frac{\partial P}{\partial \phi_\gamma} \frac{\partial P}{\partial \phi_\delta} \Bigg)\,,
\end{align}
where $\epsilon_{\alpha\beta\gamma\delta}$ is the antisymmetric tensor of rank-$4$.

Formally, the $\Snd$ Chern number is defined by $\mathcal{V} = \int d^4 \phi_\mu\;\mathcal{C}(\phi_\mu)$~\cite{Avron1989}. By its definition,
$\mathcal{V}$ characterizes a 4D family of Hamiltonians composed of all $H(\phi_\mu)$ with all possible values of $\phi_\mu$;
i.e.~$\phi_x$ and $\phi_y$ act as two additional effective dimensions. The main result of this Letter is that even for a given
$\phi_\mu$, the gaps of $H(\phi_\mu)$ can be associated with a nontrivial integer $\Snd$ Chern number,
\begin{align}     \label{Eq:second_Chern}
    \mathcal{V}=(2\pi)^4 \mathcal{C}(\phi_\mu) \neq 0 \,,
\end{align}
in the thermodynamic limit. Below, we justify Eq.~\eqref{Eq:second_Chern} by showing that, for our model, $\mathcal{C}(\phi_\mu)$
is essentially independent of $\phi_\mu$, and thus the integration over the four parameters is redundant.

Beforehand, we present the physical implications of the nontrivial $\mathcal{V}$. To do so, we apply to $H$ the procedure of
\textit{dimensional extension} that was introduced in Refs.~\cite{us,us2}. In this procedure, we interpret $\phi_x$ and $\phi_y$
as momenta along two fictitious perpendicular coordinates $w$ and $z$, respectively. Now, the Hamiltonian $H(\phi_x,\phi_y)$ is a
single Fourier component of some ancestor 4D Hamiltonian $\H$. By making the inverse Fourier transform, we obtain a Hamiltonian
describing spin-$\frac{1}{2}$ particles hopping on a 4D hyper-cubic lattice
\begin{align} \label{Eq:H4D}
    \H &= \sum_{\xvec,\hat{\mu}} \mathbf{c}_{\xvec}^\dag e^{i 2\pi a_{\hat{\mu}}(\xvec)}
          t_{\hat{\mu}} \mathbf{c}_{\xvec+\hat{\mu}} + \text{H.c.}\, ,
\end{align}
where $\mathbf{c}_{\xvec}=(c_{\xvec,+}, c_{\xvec,-})$ annihilates a spin-$\frac{1}{2}$ particle at site $\xvec = (x,y,z,w)$,
$\hat{\mu}$ is summed over the unit vectors $\hat{x},\hat{y},\hat{z}$ and $\hat{w}$, which connect nearest neighbors, and
$t_{\hat{\mu}} = (t_x,t_y,\lambda_x/2,\lambda_y/2)$. These particles are coupled to a Yang-Mills gauge field
$a_{\hat{\mu}}(\xvec) = ( b_y y \hat{z} + b_x x \hat{w} )\sigma_3$. This vector potential describes a spin-polarized uniform
SU(2) field. Such a field in known to generate a 4D IQHE with a nontrivial $\mathcal{V}$~\cite{Yang78,IQHE4D}. Notably, $\H$ is
defined on a planar geometry in a Landau gauge, whereas previous analyses treated a spherical geometry in a symmetric gauge.

Similar to the 2D IQHE, $\mathcal{V}$ has a physical manifestation in the form of a response function. Here the response is
quantized but non-linear: $j_\alpha = \mathcal{V} \frac{e^2}{h\Phi_0} \epsilon_{\alpha\beta\gamma\delta} B_{\beta\gamma}
E_\delta$~\cite{QiZhang}, where $j_\alpha$ denotes the current density along the $\alpha$ direction, $\Phi_0$ is the flux
quantum, $E_\delta$ is an electric field along the $\delta$ direction, and $B_{\beta\gamma}$ is a magnetic field in the
$\beta\gamma$ plane.

A direct observation of this response requires a 4D system. However, we can develop an analogue of Laughlin's pumping, which is
manifested in the 2D model. Let us consider the following two cases: $j_x=\mathcal{V}\frac{e^2}{h\Phi_0} B_{yz} E_w$ and
$j_x=\mathcal{V}\frac{e^2}{h\Phi_0} B_{wy} E_z$. Recall that the electric fields can be generated by time-dependent Aharonov-Bohm fluxes, $E_w = \frac{1}{caN_w} \partial_t \Phi_w(t)$ and $E_z = \frac{1}{caN_z} \partial_t
\Phi_z(t)$, where $a$ is the lattice spacing, and $N_w$ and $N_z$ are the number of lattice sites along the $w$  and
$z$ directions, respectively. Expressing $B_{yz}$ and $B_{wy}$ in $\H$ in Landau gauge, and performing dimensional reduction,
these fields enter the 2D model, $H$, through modified on-site terms,
\begin{multline} \label{Eq:modification}
\lambda_x \cos[2\pi(\sigma b_x x + \bar{B}_{wy} y) + \phi_x(t)] \\
+ \lambda_y \cos[2\pi(\sigma b_y+\bar{B}_{yz}) y + \phi_y(t)] \,,
\end{multline}
where $\phi_x(t) = \frac{2\pi}{N_w\Phi_0}\Phi_w(t)$ and $\phi_y(t) = \frac{2\pi}{N_z\Phi_0}\Phi_z(t)$, and $\bar{B}_{wy} =
B_{wy}a^2/\Phi_0$ and $\bar{B}_{yz} = B_{yz}a^2/\Phi_0$ denote the corresponding flux quanta per plaquette. Figures
\ref{system_experiment}(b)-(d) illustrate the effects of these modifications. By fixing the chemical potential within a gap
with a given $\mathcal{V}$, an adiabatic scan of $\phi_x$ or of $\phi_y$ from $0$ to $2\pi$ pumps charge along the $x$ direction,
such that
\begin{align}
Q_{x} = \mathcal{V} e\bar{B}_{yz}N_y\,,\,\,\,{\rm (6a)}\quad\quad\quad Q_{x} = \mathcal{V} e\bar{B}_{wy}N_y\,,\,\,\, {\rm (6b)}
\nonumber
\end{align}
respectively, where $N_y$ is the number of lattice sites along the $y$ direction.

We can now propose experiments that measure $\mathcal{V}$ using Eq.~(6). Take a 2D slab of our model, and connect metal leads to
the edges of the $x$ coordinate. Let us assume that the chemical potential of both the system and the leads is placed in some gap
of $H$. Then, one should measure the charge-flow during the scan of $\phi_x$ from $0$ to $2\pi$ for different values of
$\bar{B}_{yz}$ \cite{small_B}. According to Eq.~(6a), we expect that $\mathcal{V} = \frac{1}{eL_y}\partial Q_x/\partial
\bar{B}_{yz}$. Similarly, according to Eq.~(6b), during the scan of $\phi_y$ while varying $\bar{B}_{wy}$, charge flows and
$\mathcal{V} = \frac{1}{eL_y}\partial Q_x/\partial \bar{B}_{wy}$. Remarkably, due to the 4D origin of our model, the measured
$\mathcal{V}$ would be the same in both experiments. A similar experiment can be conducted in a photonic system \cite{supmat}.

\begin{figure}[tbh]
\begin{center}
\includegraphics[width=\columnwidth]{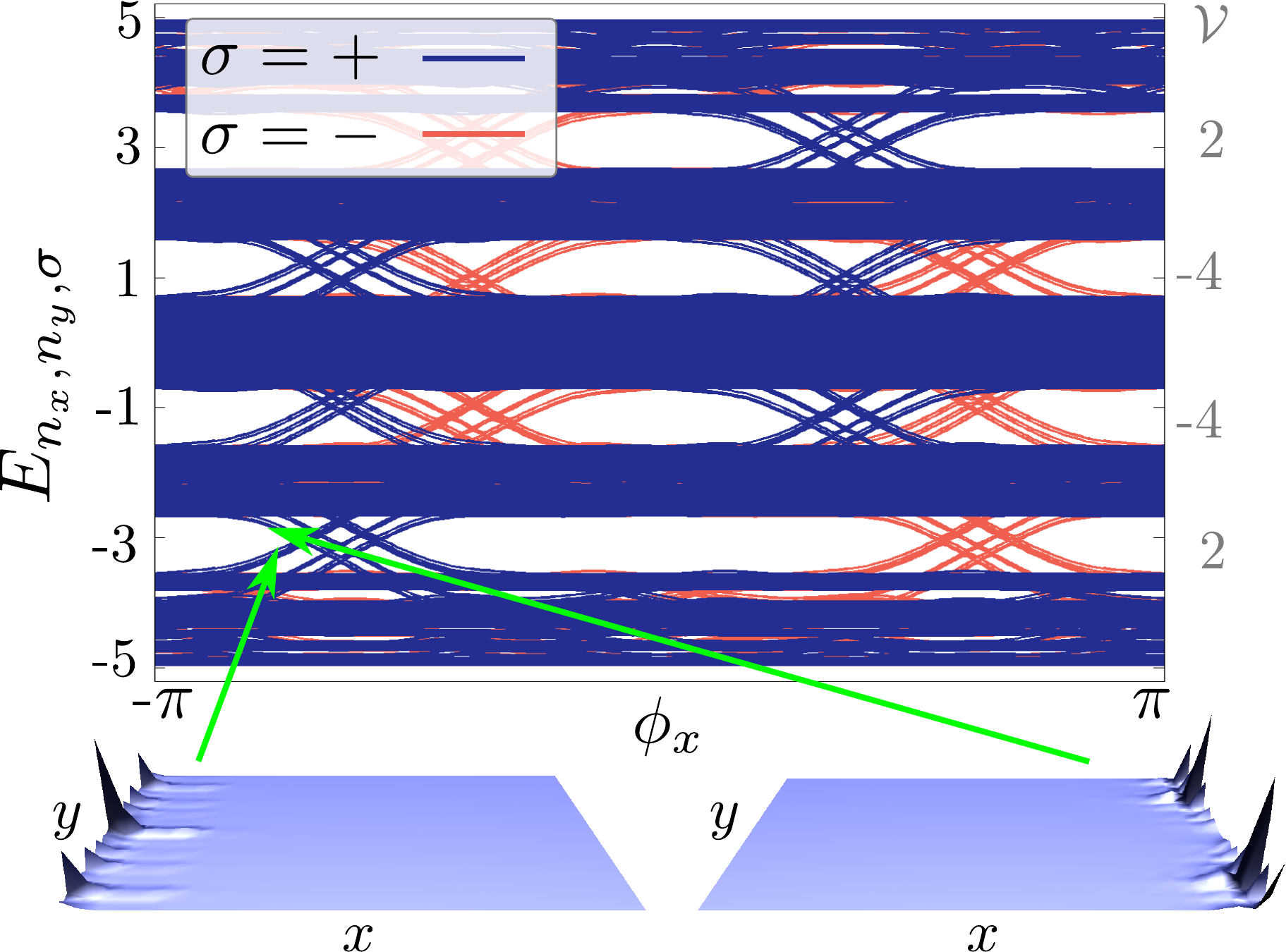}
\end{center}
\caption{  \label{spectrum} %
The spectrum of $H$ as a function of the shift parameter $\phi_x$ [cf.~Eq.~\eqref{Eq:H2D}], for an open $x$ coordinate and a
periodic $y$ coordinate. The horizontal bands correspond to bulk states. The gap-traversing states with $\sigma=+$ (blue) or
$\sigma=-$ (red) are edge states, which are localized at the left or right edges (see insets for typical wave functions). The
crossings of edge states are topologically protected. The values of the $\Snd$ Chern number, $\mathcal{V}$,
associated with the large gaps are presented.}
\end{figure}

We have just seen that upon a scan of $\phi_x$, charge may flow in the $x$ direction. Consequently, for an open geometry, in
order to accommodate this charge transfer, edge states must appear and traverse the gaps as a function of $\phi_x$. These states
appear for infinitesimally small $\bar{B}_{yz}$, and hence appear also for $\bar{B}_{yz}=0$. Figure \ref{spectrum} depicts the
numerically obtained energy spectrum of $H$ as a function of $\phi_x$, for an open $x$ coordinate, a periodic $y$ coordinate,
$t_x=t_y=1$, $\l_x=\l_y=1.8$, $N_x=N_y=34$, $b_x=(1+\sqrt{5})/2$ and $b_y=55/34 \approx (1+\sqrt{5})/2$~\cite{torus}. The spectrum
is invariant with respect to $\phi_y$, and is depicted for $\phi_y = 0$. As a function of $\phi_x$, the spectrum has flat bands
and gap-traversing bands. The flat bands correspond to bulk states, whereas the gap-traversing ones to edge state (see insets).
The edge states are divided into four types: $\sigma=+$ and $\sigma=-$ (blue and red), which are localized at either the left or
right edge (opposite slopes). These edge states are a signature of the nontrivial $\mathcal{V}$ of our model. They can be
measured in a way similar to the experiments performed in 1D photonic QCs~\cite{us,supmat}.

Naively, opposite-$\sigma$ modes that reside on the same edge can be gapped out by introducing $\sigma$-mixing terms. However, this is not the case, and the edge modes and their crossings are topologically protected. In order to
establish this protection, we decompose $H$ into its $\sigma$ and spatial constituents. The Hamiltonian $H$ is a sum of two
$\sigma$ components, where each $\sigma$ component is subject to two decoupled 1D Harper models along the $x$ and
$y$ directions. Both $\sigma$ components experience the same modulation frequencies, $b_x$ and $b_y$, but couple to the shift
parameters, $\phi_x$ and $\phi_y$, with an opposite sign. Therefore, each eigenstate of $H(\phi_x,\phi_y)$ is a product of
eigenstates of the Harper models in the $x$ and $y$ directions and an eigenstate of $\sigma$.

Recall that each gap of the Harper model is associated with a nontrivial $\Fst$ Chern number, which corresponds to the number of boundary states
that traverse the gap as a function of $\phi$~\cite{TKNN,Avron1,us}. Accordingly, each band of the Harper models in the $x$ and
$y$ directions is associated with some $\sigma$-dependent $\Fst$ Chern number, $\nu_{r_x,\sigma}$ and $\nu_{r_y,\sigma}$, where $r_x$ and $r_y$
denote the corresponding bands, respectively~\cite{mind_the_band}. In Fig.~\ref{spectrum}, the bands that traverse the gaps as a
function of $\phi_x$ are composed of products of bulk bands in the $y$ direction and boundary states in the $x$ direction.
Notably, due to the opposite coupling of $\sigma$ to $\phi_x$ and $\phi_y$, $\nu_{r_x,-} = -\nu_{r_x,+}$ and $\nu_{r_y,-} =
-\nu_{r_y,+}$. Therefore, the gaps are traversed by the same number of $\sigma=\pm$ bands, but with opposite slopes. Since
opposite-$\sigma$ bands are associated with opposite $\Fst$ Chern numbers, they cannot be gapped out by $\sigma$-mixing terms, even if they cross
at some value of $\phi_x$. Otherwise, $\nu_{r_y,\sigma}$ would change continuously as a function of $\phi_x$ between
$\nu_{r_y,+}$ and $\nu_{r_y,-}$. The level crossing is therefore protected.

The described edge phenomena accounts for the charge pumping described above. When $\phi_x$ or $\phi_y$ are scanned,
opposite-$\sigma$ states, which have opposite $\Fst$ Chern numbers, flow in opposite directions. In the absence of $\bar{B}_{yz}$ and
$\bar{B}_{wy}$, the two charge currents cancel. Applying $\bar{B}_{yz}$ or $\bar{B}_{wy}$ causes a difference between the
densities of the $\sigma$ states, and thus a net charge is pumped~\cite{QiZhang,RahulRoy}. Remarkably, for a spin-$\frac{1}{2}$
realization, our model can serve as a spin pump across the sample, since for recursive scans of $\phi_x$, macroscopic spins
accumulate at the boundaries, even for vanishing $\bar{B}_{yz}$ and $\bar{B}_{wy}$.

We turn, now, to establish Eq.~\eqref{Eq:second_Chern}. Let us evoke the definition of the $\Fst$ Chern number of a band of the Harper model in
the $x$ direction with a given $\sigma$, $\nu_{r_x,\sigma}=\int d\phi_x d\theta_x \; C_{r_x,\sigma} (\phi_x,\theta_x)$, where
$C_{r_x,\sigma}(\phi_x,\theta_x) = \frac{1}{2\pi i} \Tr (P_{r_x,\sigma}[\partial_{\phi_x} P_{r_x,\sigma},\partial_{\theta_x}
P_{r_x,\sigma}])$, and $P_{r_x,\sigma}(\phi_x,\theta_x)$ is the projection matrix on the eigenstates of the $r_x$th
band~\cite{Avron1989}. A similar definition applies for $\nu_{r_y,\sigma}$. Let us denote by $\epsilon_n(\phi)$ the eigenenergies
of the Harper model. The decomposition of $H$ into $x$  and $y$ constituents makes its energy spectrum a Minkowski sum,
$E_{n_x,n_y,\sigma}(\phi_x,\phi_y) = \epsilon_{n_x}(\sigma \phi_x) + \epsilon_{n_y}(\sigma \phi_y)$. Accordingly, the states
below each gap of $E_{n_x,n_y,\sigma}$ can be decomposed into a sum over pairs of bands in the 1D spectra, $r_x$ and $r_y$, such
that $\epsilon_{r_x} + \epsilon_{r_y} < \mu$, where $\mu$ is the energy in the middle of the gap. Using the fact that the
eigenfunctions of $H$ are a product of the 1D eigenfunctions, we obtain~\cite{supmat}
\begin{align} \label{Eq:nu2_nu1}
     \mathcal{V} = \sum_{\sigma = \pm} \sum_{(\epsilon_{r_x}+\epsilon_{r_y} < \mu)}
     \nu_{r_x,\sigma}\nu_{r_y,\sigma} \neq 0\,.
\end{align}
In a previous work~\cite{us}, we have shown that, in the thermodynamic limit, $C_{r,\sigma}(\phi,\theta)$ becomes independent of $\phi$ and
$\theta$. Hence $\nu_{r,\sigma}= (2\pi)^2 C_{r,\sigma}(\phi,\theta)$. This, combined with Eq.~\eqref{Eq:nu2_nu1}, immediately implies
Eq.~\eqref{Eq:second_Chern}~\cite{supmat}.

Until now, our analysis used the decomposition of $H$ into $\sigma$, $x$ and $y$ components. In fact, any SU(2)$\times$U(1) gauge
transformation in the 4D Hamiltonian $\H$ that respects Landau gauge keeps the system unchanged. After the dimensional reduction to 2D, such a
transformation becomes a general local transformation that may mix $x$, $y$, and $\sigma$, but keeps $\mathcal{V}$ unchanged.
More generally, any unitary transformation of the 2D Hamiltonian that does not depend on $\phi_\mu$ keeps $\mathcal{C}(\phi_\mu)$
independent of $\phi_\mu$. We also show that the symmetry of $H$ to spin flip is unnecessary~\cite{supmat}. Additionally, in the presence of uncorrelated disorder that does not close the bulk gap, $\mathcal{V}$ remains the same, similar to the 1D case~\cite{us}.

The predicted $\Snd$ Chern number is not limited to a 2D model composed of two Harper models. One can consider (i) placing the cosine modulations
in the hopping terms, rather than in the on-site terms (off-diagonal Harper), and (ii) replacing each of the cosine modulations
with a Fibonacci-like modulation. Combining (i) and (ii) leads to a well-known 2D QC \cite{Lifshitz:2008}. In 1D, all these
variants were shown to be topologically equivalent to the Harper model, namely they have the same distribution of
$\Fst$ Chern numbers~\cite{us2}. Therefore, the gaps in such 2D QCs variants will have nontrivial $\mathcal{V}$. Note, also, that similar models
that do not depend on $\sigma$ can have gaps with nontrivial $\mathcal{V}$ with accompanying bulk response and edge
phenomena~\cite{supmat}.

To conclude, in this Letter we have presented a novel 2D quasiperiodic model that is associated with the same topological index
as the 4D IQHE -- the $\Snd$ Chern number. Correspondingly, our model exhibits an elaborate edge phenomena and a quantized
charge-pump. We propose experiments in which charge is pumped through the system following modifications of the
quasiperiodic modulation. Interestingly, while these modifications differ considerably, they lead to the same pumped charge. This
equivalence may seem baffling from a 2D perspective, but follows from the symmetry of the non-linear response
in 4D. Recent progress in controlling and engineering systems, such as optical lattices~\cite{atala2012}, photonic
crystals~\cite{rechtsman2012,Hafezi:2013}, and molecule assembly on metal surfaces~\cite{Manoharan:2012}, makes our 2D model seem
experimentally feasible. Moreover, interactions in such systems may lead to fractional quasicrystalline phases which are
descendants of the exotic 4D fractional quantum Hall effect~\cite{IQHE4D}. Thus, our model serves as a porthole by which to
access 4D physics.

We thank E.~Berg and Y.~Lahini for fruitful discussions. %
We acknowledge the Minerva Foundation of the DFG, the US-Israel BSF, and the Swiss National Foundation for financial support. %
Authors' names appear in alphabetical order.


\begin{thebibliography}{39}
\bibitem[{tor()}]{torus}
\bibinfo{note}{For periodic boundary conditions, $b_x$ and $b_y$ are
  approximated by rational numbers. The resutling discrepancy vanishes in the
  thermodynamic limit.}
\bibitem[{sma()}]{small_B}
\bibinfo{note}{Notably, $\bar{B}_{yz}$ must be sufficiently small to keep the
  gap open, and should be picked such that $\bar{B}_{yz}N_y$ is approximately
  an integer.}
\bibitem[{min()}]{mind_the_band}
\bibinfo{note}{The specturm of the quasiperiodic Harper model is fractal. Its
  energy bands are defined by progressively taking rational approximants of
  $b$. The $1^\textrm{st}$ Chern number associated with a band of an
  approximant as a function of $\phi_y$ remains the same in the irrational
  limit even for a given $\phi_y$. In our model, the approximants must be
  sufficiently good such that Harper bands within a gap are narrower than the
  gap they cross. This way we can also associate each band with its mean
  energy.}
\end{thebibliography}

\newpage
\cleardoublepage
\newpage

\begin{center}
\textbf{\large SUPPLEMENTAL MATERIAL}
\end{center}
\vspace{5mm}

\setcounter{enumi}{1}
\setcounter{equation}{0}
\renewcommand{\theequation}{\Roman{enumi}.\arabic{equation}}

\section{I. Proposal for photonic experiments}

In the main text, we present a 2D quasiperiodic model [cf.~Eq.~(1)] that is characterized by the 4D $\Snd$ Chern number. It is
argued that 4D physics is manifested in the appearance of localized edge states as a function of the shift parameter
$\phi_x$, and by quantized pumping of charge. We also propose electronic experiments to measure these phenomena.
Here, we provide an elaboration on how to realize our theoretical predictions using photonic quasicrystals, following
Refs.~\cite{us,us4}.

We suggest using photonic quasicrystals that are composed of an array of coupled two-mode waveguides~\cite{Waveguides_Review2}. The
effective refractive index of each waveguide determines the phase that light accumulates during the propagation along it. The
overlap between the evanescent modes of the waveguides allows the light to tunnel from a waveguide to its neighboring waveguides,
as it propagates. The hopping rate is determined by the spacing between neighboring waveguides. Such systems have a high level of
control over their parameters and behavior.

The dynamics of light propagating along the array can be described by the tight-binding model of the Schre\"{o}dinger
equation, with the propagation axis, $z$, taking over the role of time: $i\partial_{z}\psi_{n,m,\sigma}=H\psi_{n,m,\sigma}$. Here,
$\psi_{n,m,\sigma}$ is the wavefunction at waveguide $(n,m)$ of mode $\sigma = \pm$. Note that the two $\sigma$-modes play a
role similar to a spin-$\frac{1}{2}$ internal degree of freedom. By choosing the appropriate refractive indices and inter-waveguide
spacings, one can generate an array that emulates the dynamics of our Hamiltonian [cf.~Eq.~(1) in the main text]. Notably, as we discuss below (Sec.~III of this
Supplement), one can perform, in principle, the following experiments and obtain similar results also for single-mode waveguides,
namely, for a model without $\sigma$. We, however, discuss here the full Hamiltonian, and assume that the two modes
correspond to two polarizations.

\begin{figure*}
\includegraphics[width=\textwidth,clip]{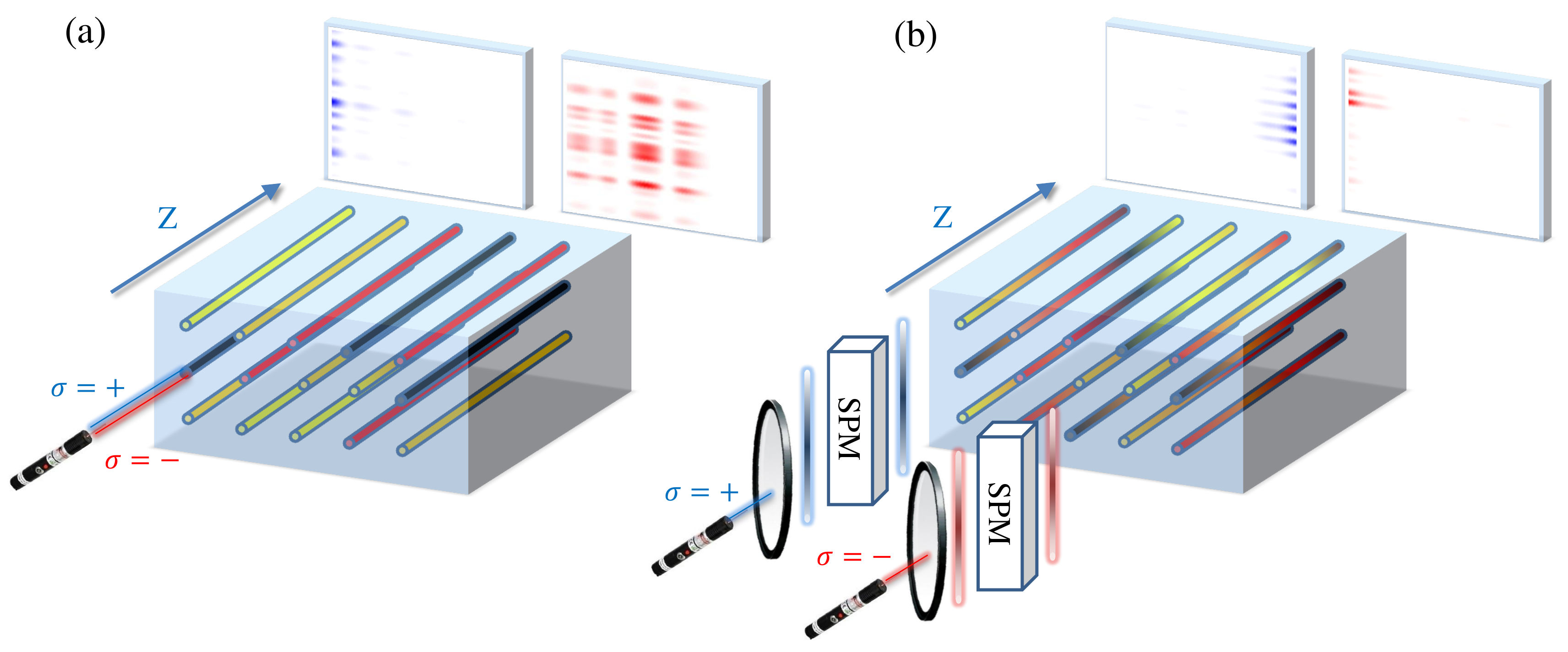}
\caption{ \label{Exp} %
Illustration of the proposed experiments to detect the topological properties of our model [cf.~Eq.~(1) in the main text] in
photonic quasicrystals. The system is made of parallel two-mode waveguides, with polarization-dependent refractive indices. The
dynamics of light propagating along $z$ is given by the tight binding model, which emulates our 2D model. (a) Subgap edge states
at a given shift $\phi_x$ are observed by injecting a polarized beam at the $x$ edge and measuring its spread across the
$x$ direction along the propagation. A beam that remains tight to the edge indicates an edge state. Here illustrated, is the
behavior of an array with edge states only for $\sigma = +$ polarization. Hence, in the left inset (blue), the $\sigma = +$ light
remains localized at the edge, and in the right inset (red), the $\sigma = -$ polarization beam broadens. (b) Adiabatic pumping
of light manifests the $\Snd$ Chern number, $\mathcal{V}$. The shift parameter $\phi_x$ is adiabatically changed along $z$, such
that the edge states of the two polarization swap their positions, i.e.~, they are counter-pumped across the $x$ direction. Each
of the injected beams is designed to excite a whole subgap band using spatial phase modulators (SPMs). When
$\bar{B}_{yz}=\bar{B}_{wy}=0$, the number of states in the $\sigma = +$ and $\sigma = -$ bands is the same, and the outgoing
intensities of the two polarization are equal. The inclusion of the modulations $\bar{B}_{yz}$ and $\bar{B}_{wy}$ differs between
the pumped intensities of the $\sigma = +$ and the $\sigma = -$ states. The difference between the two outgoing powers is linear
with $\mathcal{V}\bar{B}_{yz}$ and $\mathcal{V}\bar{B}_{wy}$.}
\end{figure*}

The properties of such waveguide arrays are studied by injecting a beam of polarized light into the waveguides at the edge of the
array, and measuring the outgoing intensity. The injected beam excites eigenstates of the Hamiltonian with the same polarization
and that have a non-vanishing amplitude at the injection sites. The light propagates along the $z$ direction according to the
superposition of the excited eigenstates. For an array with a given $\phi_x$, if there are no edge states at the edge where the
light is injected, then the beam overlaps only with extended bulk states, and spreads freely throughout the array. However, if
there are localized edge states, then the beam with the corresponding polarization overlaps mainly with these states, and remains
tight to the edge. Fig.~\ref{Exp}(a) is an illustration of such an experiment. Each waveguide in the array has a
polarization-dependent refractive index, and for the specific $\phi_x$ chosen for the array, there are left-edge states only for
the $\sigma = +$ polarization. This way, one can validate the existence of the topological edge states.

To measure the $\Snd$ Chern number of the system, $\mathcal{V}$, we propose a pumping experiment similar to the one
discussed in the main text. This experiment is based on three facts: (i) as one scans $\phi_x$, the subgap
edge states approach the bulk bands, and return into the gap localized at the opposite edge (as seen in
Fig.~2 of the main text); (ii) the number of such edge states depends on the modulations $\bar{B}_{yz}$ and $\bar{B}_{wy}$;
(iii) the pumping direction and the effect of the modulations are opposite for opposite polarizations. Therefore, one should
implement an array such that $\phi_x$ adiabatically changes along $z$, where at the starting point there are edge states of
opposite polarizations at opposite edges, and at the termination point they swap positions. Thus, when injecting two oppositely
polarized beams to the opposite edges, the two beams are counter-pumped across the array.

In order to observe $\mathcal{V}$, the injected beam should excite only a particular band, for example, by having an appropriate
phase profile, using a spatial phase modulator (SPM). Changing the on-site modulation via $\bar{B}_{yz}$ and $\bar{B}_{wy}$,
changes the band density in an opposite manner for the opposite polarizations. Therefore, for one polarization the transmission
is higher than for the other. By subtracting the outgoing power of the $\sigma=-$ light from the power of the $\sigma=+$ light, a
net power is observed, which is expected to be linear in $\mathcal{V}\bar{B}_{yz}$ and $\mathcal{V}\bar{B}_{wy}$ [see Eq.~(6) of the
main text]. An illustration of such an experiment is presented in Fig.~\ref{Exp}(b). Note that, in such an optical setup, one
usually measures $\mathcal{V}$ of the largest gap, and, thus, $\mathcal{V}$ is measured only for $\nu_{r_x,\sigma} = \pm 1$.


\section{II. Factorization of the $\Snd$ Chern number}

In the main text, we notice that the Hamiltonian $H$ [cf.~Eq.~(1) of the main text] can be decomposed into Harper models in the
$x$  and $y$ directions, for both $\sigma$ species. Consequently, we infer that the $\Snd$ Chern numbers $\mathcal{V}$ of $H$ can
be expressed as multiplications of the $\Fst$ Chern numbers of the constituting Harper models [cf.~Eq.~(7) of the main text]. In
this Supplemental Material we prove this preposition.

Recall that for a given gap in the spectrum of $H(\phi_\mu)$, we define the projection matrix $P(\phi_\mu)$ on all the
eigenstates of $H(\phi_\mu)$ with energies below this gap. Similar to $H$, $P = \sum_{\sigma = \pm} P_\sigma$, where $P_\sigma$
projects on the $\sigma$ states, and thus $P_+ P_- = 0$. Additionally, we note that the states below the gap of the 2D model,
$H$, are composed of pairs of 1D bands that satisfy $\epsilon_{r_x} + \epsilon_{r_y} < \mu$, where $\epsilon_r$ is an eigenenergy
of the $r$th band of the Harper model, and $\mu$ is an energy in the middle of the 2D gap. Therefore,
\begin{align} \label{Eq:sum_P}
\resizebox{.85\hsize}{!}{$\displaystyle P(\phi_\mu) = \sum_{\sigma = \pm} \sum_{\left(\epsilon_{r_x} + \epsilon_{r_y} <
\mu\right)} P_{r_x,\sigma}(\phi_x,\theta_x) \otimes P_{r_y,\sigma}(\phi_y,\theta_y) \,, $}
\end{align}
where $P_{r_x,\sigma}(\phi_x,\theta_x)$ is the projection matrix on the eigenstates of the $r_x$th band, and similarly for
$P_{r_y,\sigma}(\phi_y,\theta_y)$. Note that the elements of this sum are orthogonal projectors.

In the main text, we introduced the $\Snd$ Chern form, $\mathcal{C}(\phi_\mu)$ [cf.~Eq.~(2) of the main text], and the $\Snd$
Chern number, $\mathcal{V} = \int d^4 \phi_\mu\;\mathcal{C}(\phi_\mu)$~\cite{Nakahara}. Since $\mathcal{V}$ is a
$\mathbb{Z}$-index, a $\Snd$ Chern number can also be assigned to each element of the sum in Eq.~\eqref{Eq:sum_P}. It suffices,
therefore, to focus on a single such element, and show that its $\Snd$ Chern number is a product of its corresponding $\Fst$
Chern numbers.

Let us denote such an element by $\P(\phi_\mu) = P_{r_x,\sigma}(\phi_x,\theta_x) \otimes P_{r_y,\sigma}(\phi_y,\theta_y)$, and
its corresponding $\Snd$ Chern form by $\bar{\mathcal{C}}(\phi_\mu)$. For brevity, we denote $\partial \P / \partial\phi_\alpha$
by $\P^{[\alpha]}$, which makes
$\bar{\mathcal{C}}(\phi_\mu)=\epsilon_{\alpha\beta\gamma\delta}\Tr(\P\P^{[\alpha]}\P^{[\beta]}\P\P^{[\gamma]}\P^{[\delta]})/(-8\pi^2)$.
According to the decomposition of $\bar{P}$ into $x$  and $y$ parts, $\bar{\mathcal{C}}$ involves two types of terms:
\begin{align} \label{Eq:two_types}
\begin{array}{ll}
\text{(i)}  & \Tr(P_{r_x,\sigma} P_{r_x,\sigma}^{[\alpha]} P_{r_x,\sigma} P_{r_x,\sigma}^{[\gamma]} P_{r_x,\sigma}
                \otimes P_{r_y,\sigma} P_{r_y,\sigma}^{[\beta]} P_{r_y,\sigma} P_{r_y,\sigma}^{[\delta]})\,, \nonumber \\
\text{(ii)} & \Tr(P_{r_x,\sigma} P_{r_x,\sigma}^{[\alpha]} P_{r_x,\sigma}^{[\beta]} P_{r_x,\sigma}
                \otimes P_{r_y,\sigma} P_{r_y,\sigma}^{[\gamma]} P_{r_y,\sigma}^{[\delta]})
\,. \nonumber
\end{array}
\end{align}
The type-(i) terms vanish, since for any projector, $P$, the following applies: $P P^{[\alpha]} P  = 0$, where $P^{[\alpha]}$ is
a derivative of $P$ with respect to some variable $\alpha$. Recall that for any two matrices, $\Tr(A \otimes B) = \Tr(A) \Tr(B)$.
This makes the type-(ii) terms become $\Tr(P_{r_x,\sigma} P_{r_x,\sigma}^{[\alpha]} P_{r_x,\sigma}^{[\beta]}) \Tr(P_{r_y,\sigma}
P_{r_y,\sigma}^{[\gamma]} P_{r_y,\sigma}^{[\delta]})$. Substituting this into $\bar{\mathcal{C}}$, we obtain
\begin{align} 
\bar{\mathcal{C}}(\phi_\mu) &= \frac{\epsilon_{\alpha\beta}}{2\pi i} \Tr( P_{r_x,\sigma} P_{r_x,\sigma}^{[\alpha]}
P_{r_x,\sigma}^{[\beta]}) \cdot \frac{\epsilon_{\gamma\delta}}{2\pi i}\Tr( P_{r_y,\sigma} P_{r_y,\sigma}^{[\gamma]}
P_{r_y,\sigma}^{[\delta]}) \nonumber \\
            &\equiv C_{r_x,\sigma}(\phi_x,\theta_x) \; C_{r_y,\sigma}(\phi_y,\theta_y) \,,
\end{align}
where $\epsilon_{\alpha\beta}$ is the antisymmetric tensor of rank $2$, and $C(\phi,\theta)$ is the $\Fst$ Chern
form~\cite{Nakahara}. The $\Snd$ Chern number corresponding to $\bar{\mathcal{C}}$ is
\begin{align}
\bar{\mathcal{V}} &= \int d^4 \phi_\mu\;\bar{\mathcal{C}}(\phi_\mu) \\
&= \int d\phi_x d\theta_x\;C_{r_x,\sigma} \cdot \int d\phi_y d\theta_y\;C_{r_y,\sigma} = \nu_{r_x,\sigma} \nu_{r_y,\sigma} \,,
\nonumber
\end{align}
where $\nu_{r_x,\sigma}$ and $\nu_{r_y,\sigma}$ are the $\Fst$ Chern numbers associated with $P_{r_x,\sigma}(\phi_x,\theta_x)$
and $P_{r_y,\sigma}(\phi_y,\theta_y)$, respectively.

The quantity $\bar{\mathcal{V}}$ denotes the $\Snd$ Chern number of a single term in the sum \eqref{Eq:sum_P}, i.e.~with given $r_x$, $r_y$ and $\sigma$. Summing over all terms, we obtain the total $\Snd$ Chern number, $\mathcal{V}$, given by Eq.~(7) of the main text.


\section{III. Breaking $\sigma$-flip symmetry}

The invariance of $C(\phi,\theta)$ of the Harper model with respect to $\phi$ follows from the correspondence of $\phi$ to lattice
translations. The modulated potential $\cos(2\pi b x + \phi)$ implies that a translation by $n$ lattice sites is equivalent
to a shift of $\phi$ by $2\pi (b n \, \text{mod} \, 1)$. Provided that $b$ is irrational, the set of $\phi$-shifts generated by
all lattice translations is dense in the interval $[0,2\pi]$, in the thermodynamic limit. Consequently, any shift of $\phi$ is
equivalent, to arbitrary precision, to some lattice translation, which has no effect on $C(\phi,\theta)$. Turning back to $H$,
the invariance of $\mathcal{C}(\phi_\mu)$ with respect to $\phi_x$ and $\phi_y$ results from the equivalence of their shifts to
lattice translations in $x$ and $y$, respectively, where the $\sigma$ components are translated in opposite directions.

Our analysis leading to Eq.~(7) of the main text used the $\sigma$-flip symmetry of $H$: $\sigma \rightarrow -\sigma$. Let us break this symmetry by making the modulation frequencies $\sigma$-dependent, i.e.~$b_x\rightarrow b_{x,\sigma}$ and $b_y\rightarrow b_{y,\sigma}$. Now,
a translation by $n$ lattice sites, for example in the $x$ direction, is equivalent to $\sigma$-dependent shifts of $\phi_x$ by $2\pi (\sigma b_{x,\sigma} n \, \text{mod} \, 1)$. Ostensibly, $\phi_x$ is no longer equivalent to lattice translations. Nonetheless, if $b_{x,+}$ and $b_{x,-}$ are mutually irrational, the Kronecker-Weyl theorem~\cite{Sinai1976} implies that the space of $\sigma$-dependent shifts generated by lattice translations along $x$ is dense over $[0,2\pi]^2$. In particular, the line created by shifts of $\phi_x$ is also densely covered. Consequently, here too, any shift of $\phi_x$ is equivalent, to arbitrary precision, to some lattice translation. The same argumentation applies in the $y$ direction. Thus, Eq.~(3) of the main text holds also in the absence of $\sigma$-flip symmetry. Notably, the above symmetry breaking represents the presence of a U(1) magnetic field in the 4D Hamiltonian $\H$.


\section{IV. The role of $\sigma$}

The model discussed in the main text [cf.~Eq.~(1)] describes particles with a spin-like internal degree of freedom, $\sigma=\pm$.
While this degree of freedom is imperative for some of our reported results, some features of the model could be realized also
in a spinless model.

Let us take only one ``spin'' component of the model, e.g.~$\sigma=+$. Following the analysis leading to Eq.~(7) (of the main
text) and Section II (of this Supplemental Material), we conclude that the $\Snd$ Chern number of each gap in this spinless system is half that of the spinful model, and in particular nonzero. Moreover, the energy spectrum of the bulk of the system is the same as in the
spinful case, only with half the density of states. Accordingly, its edge spectrum contains only one of the spin components (the blue
states in Fig.~2 of the main text). Therefore, the proposed pumping experiments [cf.~discussion of Eq.~(6) of the main text]
hold, where only the slope of the pumped charge as a function $\bar{B}_{yz}$ or $\bar{B}_{wy}$ becomes shallower, corresponding to the
halved $\Snd$ Chern number. The photonic pumping experiment (cf. Section I above) is similar, but now the $\Snd$ Chern number corresponds to the pumped intensity of the light and not to the net power. Note also, that similar to the spinful case, the $\Snd$ Chern number of a gap is robust to terms that mix the $x$  and
$y$ coordinates and to uncorrelated disorder, as long as they do not close the bulk gap.

The spinless model is simpler to analyze and probably easier to implement. However, it is somewhat less elegant, both
experimentally and theoretically. Experimentally, it does not have protected band crossings of opposite $\sigma$ bands, and for a
spin-$\frac{1}{2}$ system it cannot be used as a topological spin-pump. Theoretically, it mixes effects associated with the 2D
and 4D IQHEs. We have seen that the Hamiltonian is a sum of two independent Harper models along the $x$  and $y$ directions.
Therefore, for the spinless case, both $\nu_{r_x,+}$ and $\nu_{r_y,+}$ are nonzero. In other words, a gap in the spinless model
is associated with two, generally nonzero, $\Fst$ Chern numbers, as well as a $\Snd$ Chern number, which is equal to their
multiplication. Consequently, in the proposed experiments, charge is pumped even for $\bar{B}_{yz}=\bar{B}_{wy}=0$, according to
the $\Fst$ Chern numbers. While only the $\bar{B}$-dependent contributions are proportional to the $\Snd$ Chern number. In
contrast, in the spinful model, the $\Fst$ Chern numbers vanish, and the pumping is purely due to the non-vanishing $\Snd$ Chern
number. Mathematically speaking, the spinless model has a nontrivial Chern-character, but a trivial Chern-class; while in the
spinful model, they are equal and nontrivial~\cite{Nakahara}. Therefore, it is a matter of convention whether to call the
spinless model a 4D IQHE, or simply a combination of two independent 2D IQHEs.

\end{document}